%% file: main.tex
\documentclass{article}
\usepackage{spconf,amsmath,graphicx,hyperref}
\usepackage{amsmath,graphicx}
\usepackage{booktabs}
\usepackage{multirow}
\usepackage[table]{xcolor}
\definecolor{customblue}{HTML}{1F77B4}
\usepackage[utf8]{inputenc}
\usepackage{amsfonts}
\usepackage[spanish, english]{babel}
\usepackage{tcolorbox}
\usepackage{amsthm}  
\usepackage{mathrsfs}
\usepackage{xparse}
\usepackage{algorithm}
\usepackage{algpseudocode}
\tcbuselibrary{skins}
\usepackage{color}
\usepackage{makecell, cellspace}
\usepackage{setspace}
\usepackage{graphicx}
\usepackage{multirow}
\definecolor{Pastel}{RGB}{245,245,250}
\definecolor{Pastel2}{gray}{0.99}
\usepackage{verbatim}
\makeatletter
\newcommand*\bigcdot{\mathpalette\bigcdot@{.5}}
\newcommand*\bigcdot@[2]{\mathbin{\vcenter{\hbox{\scalebox{#2}{$\m@th#1\bullet$}}}}}
\makeatother
\usepackage{bibentry}
\usepackage{makecell, cellspace}
\usepackage{booktabs}
\usepackage[normalem]{ulem}
\usepackage{mathtools, physics}
\usepackage{hyperref}
\usepackage{cleveref} 
\crefname{section}{Sec.}{Secs.} 
\usepackage{algorithm}
\usepackage{algpseudocode}
\crefname{subsection}{Sec.}{Secs.}
\crefname{subsubsection}{Sec.}{Secs.}
\crefname{figure}{Fig.}{Figs.}
\crefname{table}{Table}{Tables}
\crefname{equation}{Eq.}{Eqs.}
\crefname{algorithm}{Alg.}{Algs.} 
\crefname{theorem}{Thm.}{Thms.}
\crefname{lemma}{Lem.}{Lems.}
\crefname{proposition}{Prop.}{Props.}
\crefname{corollary}{Cor.}{Cors.}
\crefname{definition}{Def.}{Defs.}
\makeatletter

\def\@IEEEreftext#1#2{#1}%

\makeatother
\usepackage{aliascnt}
\usepackage{caption}

\usepackage{flushend}

\usepackage{caption}
\usepackage[subrefformat=simple, labelformat=simple]{subcaption}

\usepackage{flushend}
\usepackage{arydshln} 

\newcounter{remark}

\usepackage{enumitem}

\usepackage{amsmath}

\usepackage{amsmath}
\usepackage{siunitx}
\input{notation}
\def\BibTeX{{\rm B\kern-.05em{\sc i\kern-.025em b}\kern-.08em
    T\kern-.1667em\lower.7ex\hbox{E}\kern-.125emX}}
\begin{document}
\ninept

\title{Wrapper-Aware Rate-Distortion Optimization \\ in Feature Coding for Machines}

\name{Samuel Fernández-Menduiña$^{\color{magenta}\textsc{ sc}}$, Hyomin Choi$^{\color{magenta}\textsc{ id}}$, Fabien Racapé$^{\color{magenta}\textsc{ id}}$, Eduardo Pavez$^{\color{magenta}\textsc{ sc}}$, and Antonio Ortega$^{\color{magenta}\textsc{ sc}}$}

\medskip

\address{\textit{$^{\color{magenta}\textsc{sc}}$University of Southern California, Los Angeles, CA, USA} \\
\textit{$^{\color{magenta}\textsc{id}}$InterDigital AI Lab, Los Altos, CA, USA}}

\maketitle

\begin{abstract}
Feature coding for machines (FCM) is a lossy compression paradigm for split-inference. The transmitter encodes the outputs of the first part of a neural network before sending them to the receiver for completing the inference. Practical FCM methods ``sandwich'' a traditional codec between pre- and post-processing neural networks, called wrappers, to make features easier to compress using video codecs. Since traditional codecs are non-differentiable, the wrappers are trained using a proxy codec, which is later replaced by a standard codec after training. These codecs perform rate-distortion optimization (RDO) based on the sum of squared errors (SSE). Because the RDO does not consider the post-processing wrapper, the inner codec can invest bits in preserving information that the post-processing later discards. In this paper, we modify the bit-allocation in the inner codec via a wrapper-aware weighted SSE metric. To make  wrapper-aware RDO (WA-RDO) practical for FCM, we propose: 1) temporal reuse of weights across a group of pictures and 2) fixed, architecture- and task-dependent weights trained offline. Under MPEG test conditions, our methods implemented on HEVC match the VVC-based FCM state-of-the-art, effectively bridging a codec generation gap with minimal runtime overhead relative to SSE-RDO HEVC.
\end{abstract}


\section{Introduction}
\label{sec:intro}
Much of today's multimedia content is processed by vision analytics systems based on neural networks (NNs). Since images and videos are often lossy encoded and decoded before inference, some coding methods aim to specifically mitigate the impact of compression errors on downstream task performance, a framework known as \emph{coding for machines} \cite{bajic2025rate}. In applications like object tracking and instance segmentation in autonomous driving, aerial navigation, and surveillance \cite{ascenso_jpeg_2023, jiang_adaptive_2023}, the receiver may not need a copy of the original multimedia content \cite{zhang_call_2022}. Instead, it only requires the output of the NN. Thus, inference can be distributed across devices, with different model parts running on the transmitter and the receiver \cite{kang2017neurosurgeon}. Unlike \emph{local inference}—running the whole model at the transmitter and sending only labels—or \emph{remote inference}—sending the compressed video to the receiver for inference— \emph{feature coding for machines} (FCM) splits a model into two parts, 
compressing the output of the first part before sending it to the receiver, which completes the inference.
FCM distributes the computational burden between devices, enables multiple tasks at the receiver \cite{fernandez2025image}, and discards irrelevant content via task-specific pre-processing \cite{choi_deep_2018}, which can significantly reduce the bit rate relative to remote inference.

\begin{figure}
    \centering
    \includegraphics[width=\linewidth]{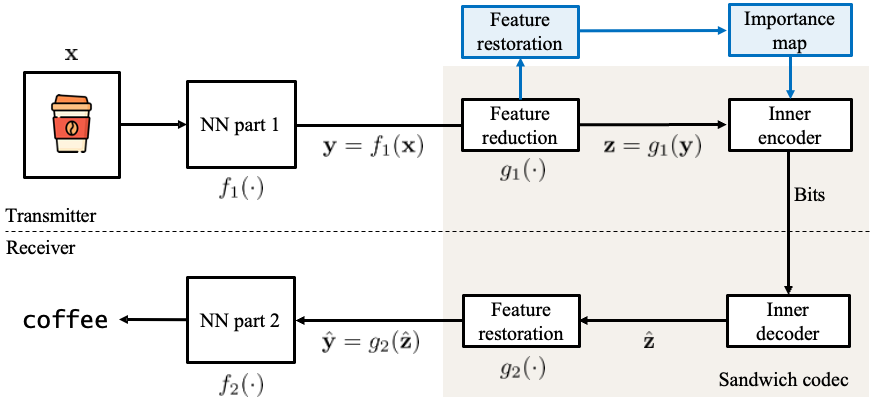}
    \caption{Sandwich-based FCM setup. The feature reduction and restoration blocks are the wrappers. Our additions appear in blue.}
    \label{fig:fcm_pipeline}
\end{figure}

Since typical image feature extractors process information localized in space \cite{krizhevsky_imagenet_2012}, features exhibit local spatial correlations, which block-wise transform coding can exploit \cite{choi_deep_2018}. Hence, practical FCM pipelines \cite{WG04_N0706_2025} build upon block-based video codecs, such as VVC \cite{bross2021overview}, due to their prevalence in hardware and software systems. Nevertheless, traditional codecs are optimized for compressing natural images, whose statistical properties differ from those of NN features. 
FCM methods mitigate this mismatch via a sandwich-based strategy \cite{guleryuz_sandwiched_2021} (\cref{fig:fcm_pipeline}): rather than compressing features directly, the inner codec is wrapped between two shallow NNs. In the transmitter, the \textit{reduction network} transforms $\bt y$, the stack of features corresponding to a video frame, into a representation more suitable for compression with traditional video codecs. Each of the feature channels at the output of the restoration network can be seen as a small 2D frame. These channels are rearranged into a larger frame, denoted by $\bt z$, where each channel corresponds to a 2D tile in the larger frame and the tiles are placed in a fixed raster-scan order \cite{WG04_N0706_2025}. In the receiver, the \textit{restoration network} aims to recover $\bt y$ from the output of the inner decoder $\hat{\bt z}$. 

Since the inner codec is non-differentiable, wrappers must be trained with a differentiable proxy codec \cite{said2022differentiable, He_2022_CVPR}, typically a learned codec \cite{kim2023end}. In particular, the wrappers are trained to minimize the sum of squared errors (SSE) between the input to the reduction network $\bt y$ and the output of the restoration network $\hat{\bt y}$, while the parameters of the pre-trained codec are kept fixed. 
For inference, this proxy is replaced by a \textit{conventional inner codec}, which performs standard bit allocation via rate-distortion optimization (RDO) to minimize the SSE between its input $\bt z$ and output $\hat{\bt z}$. 
However, minimizing SSE at the inner codec level does not imply better task accuracy, since the post-processing wrapper may discard information the inner codec encodes with high accuracy (and vice versa). 
In this paper, we use a conventional inner codec but propose a new distortion metric for the RDO that allows the bit allocation process to take into account the effect of feature compression on the task. 
Our proposed distortion metric captures the relative importance of each block in $\bt z$  on the overall error between ${\bt y}$ and $\hat{\bt y}$. Thus, we can reduce the impact of using a conventional inner codec on task performance.

To account for the restoration wrapper during encoding, we follow the Jacobian-based approach we proposed for feature-preserving RDO (FP-RDO) \cite{fernandez2024feature, fernandez2025image}. 
In our prior remote inference work \cite{fernandez2025image}, we modified a video encoder so that compression in the pixel domain preserved features in the initial layers of a remote downstream NN. Instead, in the FCM setting considered here, the codec operates on \textit{sequences of localized features}, ${\bt z}$, and the distortion is determined by the squared error between ${\bt y}$ and $\hat{\bt y}$. We show that this bit-allocation problem can be reformulated by replacing the standard SSE, i.e., 
$\|{\bt z} - \hat{\bt z}\|_2^2$, by a weighted SSE 
where weights are given by an importance map derived from the Jacobian of the restoration wrapper. 
The resulting method, termed \emph{wrapper-aware RDO} (WA-RDO), has lower computational cost than evaluating the distortion with the block-wise Jacobian formulation used in FP-RDO.

Moreover, the restoration wrapper introduces structures that can be used to simplify the bit allocation. First, we exploit temporal redundancies in the features \cite{mobahi2009deep} by computing the importance map only for intra-coded (I) frames. We then reuse the same map across the whole group of pictures (GOP). Second, given a fixed task and pair of wrappers, and regardless of the input video, the reconstruction network systematically prioritizes some of the feature channels. Hence, different spatial regions of $\bt z$ will typically receive different importance (cf.~\cref{fig:bit_allocation}). We exploit this effect to derive a characteristic importance pattern for a given wrapper by averaging the importance maps from a set of training examples. Then, we freeze the resulting pattern for RDO, instead of computing the Jacobian in an input-dependent basis. This approach removes the need to keep and evaluate the restoration network in the encoder, making our approach suitable for very low-resource devices. 

We verify our method on the common test conditions of the MPEG FCM test model (FCTM), which considers object detection, instance segmentation, and object tracking on videos and images of different resolutions and lengths.  As inner codecs, we consider AVC \cite{wiegand_overview_2003} and HEVC \cite{sullivan_overview_2012}. Results show that, using HEVC as inner codec with WA-RDO, we are able to match the performance of the current VVC-based anchor using SSE-RDO, which is the state-of-the-art \cite{compressai_vision}. Similarly, for FCTM using AVC as the inner codec, we can match the performance of FCTM-HEVC using SSE-RDO, effectively closing a codec generation gap. The temporal and architectural simplifications add small overhead over the SSE-RDO version of each codec with similar compression efficiency to WA-RDO.

\section{Preliminaries}
\noindent\textbf{Notation.} Lowercase bold letters, such as $\bt a$, denote vectors. Capital bold letters, such as $\bt A$, denote matrices. We use $\img$ to denote pixel-domain videos, $\feat$ for their features, and $\sfeat$ for the inner codec inputs. The $n$th entry of $\bt a$ is $a_n$, and the $(i, j)$th entry of $\bt A$ is $A_{ij}$.

\smallskip 

\noindent{\textbf{Sandwich codec.}} We consider the pipeline in \cref{fig:fcm_pipeline}. Let $f_1(\cdot)$ and $f_2(\cdot)$ be the transmitter (NN part 1) and receiver (NN part 2) sides of the NN, respectively. Let the wrappers be $g_1(\cdot)$ (encoder) and $g_2(\cdot)$ (decoder). We first obtain features $\feat = f_1(\img)$ and apply the encoder-side wrapper $\sfeat = g_1(\feat)$. Each channel at the output of $g_1(\cdot)$ can be seen as a small 2D frame. These are arranged in raster-scan order (left to right, top to bottom) to obtain a larger 2D frame \cite{WG04_N0706_2025}. As a result, we obtain a sequence of images, which can be compressed using the inner codec. Let $\csfeat(\thetavec)$ be the compressed version of $\bt z$ using $\thetavec\in\Theta$, where $\Theta\subset\mathbb{N}^{n_b}$ is the set of all possible parameters and $\nummb$ is the number of blocks.  We apply the post-processing part of the wrapper $\cfeat(\thetavec) = g_2(\csfeat(\thetavec))$, and finally run the second part of the NN $f_2(\cdot)$ on $\cfeat(\thetavec)$ to obtain the inference results.

\smallskip

\noindent{\textbf{SSE-based RDO.}} Given blocks of size $n_{pb}$, $\sfeat_i\in\mathbb{R}^{n_{pb}}$ for $i = 1, \hdots, \nummb$, SSE-RDO aims to find parameters $\thetavec^\star$ satisfying \cite{ortega_rate-distortion_1998}:
\begin{equation}
    \thetavec^\star = \argmin_{\thetavec \in \Theta} \, \| \csfeat(\thetavec)-\sfeat\|_2^2 + \lambda \, \sum_{i = 1}^{n_b}\, r_i(\csfeat_i(\thetavec)),
\end{equation}
where $r_i(\cdot)$ is the rate for the $i$th coding unit, and $\lambda\geq 0$ is the Lagrangian controlling the RD trade-off. Since the SSE decomposes as the sum of block-wise SSEs, $\| \csfeat(\thetavec) - \sfeat \|_2^2 = \sum_{i = 1}^{\nummb} \, \| \csfeat_i(\thetavec) - \sfeat_i \|_2^2$, and each coding unit can be optimized independently, we obtain $\csfeat_i(\thetavec) = \csfeat_i(\theta_i)$ \cite{ortega_rate-distortion_1998, fernandez2025rate}, which leads to 
\begin{equation}
\label{eq:final_form}
  \theta_i^\star = \argmin_{\theta_i \in \Theta_i} \, \|  \csfeat_i(\theta_i) - \sfeat_i\|_2^2 + \lambda \, r_{i}(\csfeat_i(\theta_i)),  
\end{equation}
for $i = 1, \hdots, \nummb,$ where $\Theta_i$ is the set of all  parameters for the $i$th block. This is the RDO formulation most video codecs solve \cite{ortega_rate-distortion_1998}. 

\section{Wrapper-aware RDO}


\noindent{\textbf{Problem statement.}}
The SSE-RDO of \eqref{eq:final_form} may be inefficient for FCM since it does not consider the post-processing wrapper. To adapt the standard inner codec to the restoration wrapper, we propose a problem formulation that seeks to minimize the difference between $\cfeat(\thetavec) = g_2(\csfeat(\thetavec))$ and  $\feat$, subject to a bit rate constraint:
\begin{equation}
\label{eq:featpres}
\thetavec^\star = \argmin_{\thetavec \in \Theta} \, \norm{ \feat -  g_2(\csfeat(\thetavec))}_2^2 + \lambda \, \sum_{i = 1}^{n_b}\, r_i(\csfeat_i(\thetavec)),
\end{equation}
We approximate the distortion term in \eqref{eq:featpres} by a weighted SSE involving the sketched Jacobian of $g_2(\cdot)$, i.e., the restoration network. Since distributed scenarios often constrain encoder complexity, we propose two simplifications to the Jacobian computation pipeline. The next paragraphs detail each of these steps.

\begin{figure}
    \centering
    \includegraphics[width=\linewidth]{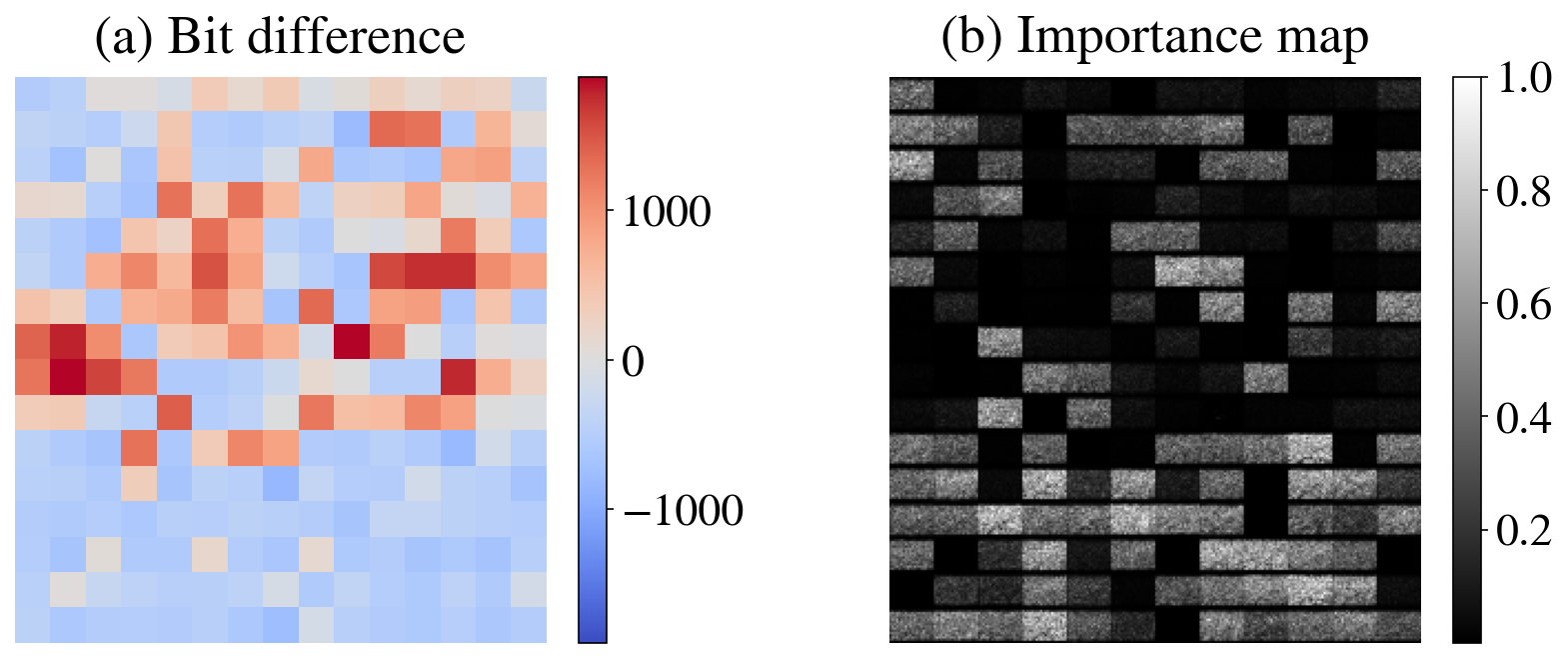}
    \caption{(a) Bit difference per macroblock using AVC with SSE-RDO and WA-RDO (feature channels arranged as an image, larger values mean SSE-RDO invests more bits than WA-RDO), and (b) WA-RDO importance map, defined for each entry of the features to compress. Since some feature channels have low importance for the restoration wrapper, WA-RDO invests fewer bits to reconstruct them.}
    \label{fig:bit_allocation}
\end{figure}

\smallskip

\noindent{\textbf{Jacobian approximation.}} 
Let $\pmb{\eta} \doteq \feat - g_2(\sfeat)$ and $\bt e(\thetavec) \doteq g_2(\csfeat(\thetavec)) -  g_2(\sfeat)$.  Expanding the distortion term in \eqref{eq:featpres}:
\begin{equation}
\label{eq:distortion}
\norm{ \feat -  g_2(\csfeat(\thetavec))}_2^2 = \| \bt e(\thetavec)\|_2^2 + \| \pmb{\eta}\|_2^2 - 2\, \langle \bt e(\thetavec), \pmb{\eta} \rangle.
\end{equation}
We discard $\norm{\pmb{\eta}}_2^2$ since it is independent of $\thetavec$. The last term in \eqref{eq:distortion} measures the correlation of the compression error at the output of the restoration wrapper with the approximation error $\pmb{\eta}$. Since FCM focuses on the almost lossless regime \cite{compressai_vision}, we can apply a high bit rate assumption, writing $\bt e(\thetavec) \cong \bt J_g(\sfeat)(\csfeat(\thetavec) - \sfeat)$ \cite{linder_high-resolution_1999}, where $\bt J_g(\sfeat)$ is the Jacobian matrix of the restoration wrapper $g_2(\cdot)$ evaluated at $\sfeat$. In the high bit rate regime,  the quantization noise can be assumed to be white \cite{gish1968asymptotically} and $1/n \, \langle \bt J_g(\sfeat)(\csfeat(\thetavec)-\sfeat), \pmb{\eta} \rangle \to 0$ as $n$ increases. Thus, we assume the first term in \eqref{eq:distortion} dominates the error, which is further reinforced when the wrapper is trained to be approximately idempotent, i.e., $g_2(g_1(\feat)) \approx \feat$. We validate this assumption in \cref{sec:exper}. As with other high-rate approximations, this assumption may degrade at very low bitrates; however, our results show that the approximation remains effective in the operating range of the MPEG FCTM common test conditions. Then, \eqref{eq:featpres} becomes
\begin{equation}
\label{eq:og_problem}
    \thetavec^\star = \argmin_{\thetavec \in \Theta} \, \norm{ \bt J_g(\sfeat)(\csfeat(\thetavec)-\sfeat)}_2^2 + \lambda \, \sum_{i = 1}^{n_b}\, r_i\big(\csfeat_i(\thetavec)\big).
\end{equation}
When $\bt y\in\mathbb{R}^{n_{\mathsf{f}}}$, the Jacobian has dimension $n_{\mathsf{f}}\times n_p$, which makes its exact computation unfeasible in practical systems. Following our prior work \cite{fernandez2025image}, we can address this problem by sketching the Jacobian with a wide random matrix $\bt S\in\mathbb{R}^{n_s\times n_p}$, with $n_p$ the number of pixels and $n_s$ the number of sketching samples. As a result, we obtain $\jacos(\sfeat) \doteq \bt S\bt J_g(\sfeat)$. By the Johnson–Lindenstrauss lemma, we can guarantee that $\norm{ \bt \jacos(\sfeat)(\csfeat(\thetavec)-\sfeat)}_2^2$ will be within $\epsilon$ of $\norm{ \bt J_g(\sfeat)(\csfeat(\thetavec)-\sfeat)}_2^2$, with $\epsilon$ controlled by $n_s$ \cite{fernandez2025image}.

\smallskip

\noindent\textbf{From Jacobian to importance map.} Let $\bt h(\sfeat)$ be the diagonal of $\bt H_s(\sfeat) = \jacos(\sfeat)^\top \jacos(\sfeat)$, which is the sketched version of the Hessian matrix of the distortion term $\| g_2(\sfeat) - g_2(\csfeat(\thetavec))\|_2^2$. The vector $\bt h(\sfeat)$ can be seen as an importance map \cite{fernandez2025image} (\cref{fig:bit_allocation}b): its $i$th entry reflects the importance of the $i$th entry of $\sfeat$ on the overall distortion. 
We use this importance map directly for bit-allocation. First, let the quantization error be $\bt q(\thetavec) = \csfeat(\thetavec) - \sfeat$. At high bit-rates, the entries of $\bt q(\thetavec)$ are uncorrelated, so $1 / n_p \, \bt q(\thetavec)^\top \bt H_s(\sfeat)\bt q(\thetavec) \to 1/n_p\, \bt q(\thetavec)^\top \diag{ h(\sfeat)}\bt q(\thetavec)$,
almost surely as $n_p$ increases. By adding SSE regularization \cite{fernandez2025image}, and for $i = 1, \hdots, n_b$,
\begin{multline}
\label{eq:wa-rdo}
    \theta_i^\star = \argmin_{\theta_i\in\Theta_i} \, \l \csfeat_i(\thetavec) - \sfeat_i \r^\top \diag{ h_i(\sfeat)}(\csfeat_i(\thetavec) - \sfeat_i) \\ + \tau \norm{\csfeat_i(\thetavec) - \sfeat_i}_2^2
    + \lambda \, r_i(\csfeat_{i}(\theta_i)).
\end{multline}
This formulation, called wrapper-aware RDO (WA-RDO), reduces memory and computational complexity by a factor of $\numfeat/n_s$ relative to \eqref{eq:og_problem}. Following \cite{fernandez2025rate}, we split $\tau = \alpha\tilde{\tau}$, with $\alpha$ balancing the SSE trade-off. Moreover, $\tilde{\tau} = \norm{\bt h(\sfeat)}_2$, and $\lambda =   \, \big( \|\bt h(\sfeat)\|_2/n_{p} + \tau\big)\,\lambda_\mathrm{SSE}$, where $\lambda_\mathrm{SSE}$ is the SSE-RDO Lagrangian \cite{ringis_disparity_2023}.

\begin{figure}
    \centering
    \includegraphics[width=\linewidth]{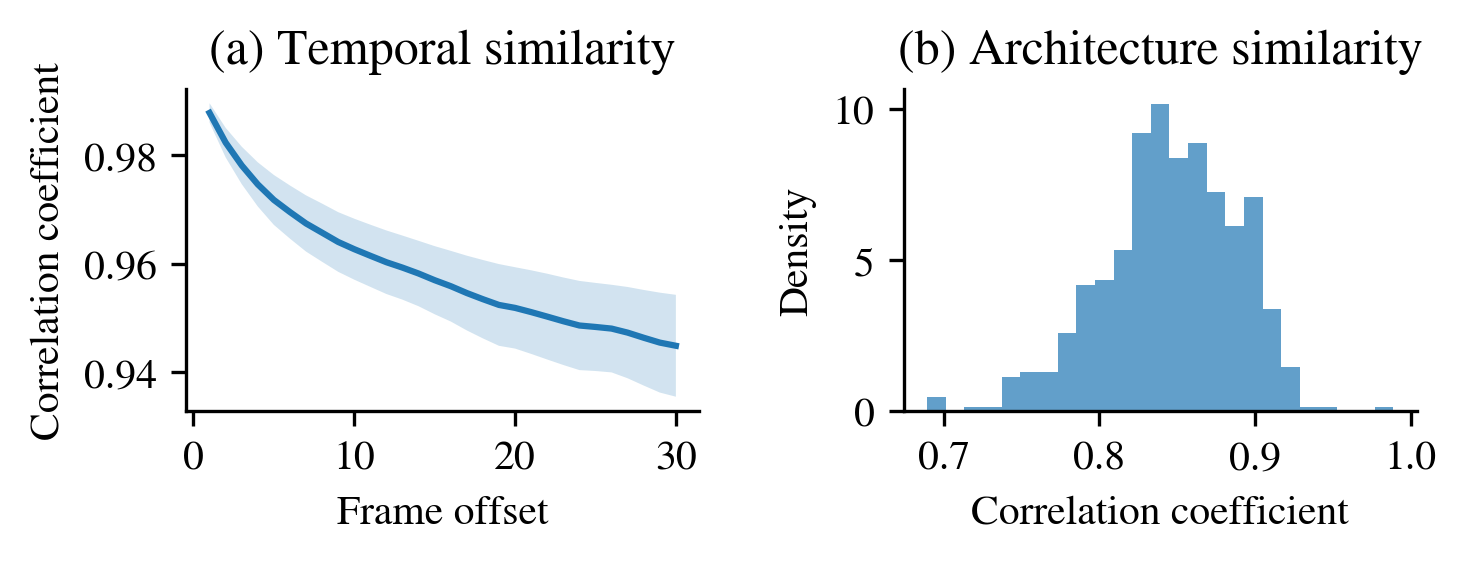}
    \caption{Correlation coefficient (CC) between importance maps. (a) Average CC across the SFU dataset as a function of the separation to the I frame with $95\%$ confidence intervals, and (b) distribution of the CC for a subset of the pairs of I frames in SFU and images in MPEG-OIV6. We observe temporal and architectural consistencies. }
    \label{fig:correlation}
\end{figure}

\smallskip

\noindent\textbf{Temporal simplification.} Computing an importance map for each video frame is computationally impractical,  requiring a forward pass of the restoration wrapper and multiple (as many as the sketching dimension) backward passes \cite{fernandez2025image}. Nonetheless, temporal redundancy in the video results in temporal redundancy in the features \cite{mobahi2009deep}. Thus, we can exploit the temporal consistency of the importance maps (cf.~\cref{fig:correlation}a) and, instead of computing the importance map for each frame, we compute the importance map for each I frame, and then reuse it for the P and B frames in between. 
This method, called \emph{I-frame Wrapper-Aware RDO (IWA-RDO)}, reduces computational complexity at the expense of adaptability to the content.

\smallskip

\noindent\textbf{Architectural simplification.} WA-RDO and IWA-RDO require encoder-side access to the restoration wrapper, which is a decoder-side component. However, we observe (\cref{fig:bit_allocation}) that the importance map depends on: 1) input-dependent scene variations, and 2) architecture-dependent variations, i.e., the wrapper systematically prioritizes some feature channels over others (\cref{fig:correlation}b). 
Since feature channels are arranged spatially in raster-scan order before compression, we can identify regions of $\bt z$ that will be given less importance by the post-processing wrapper, regardless of the input video. We exploit this to reduce complexity by fixing the restoration wrapper and marginalizing over the inputs, obtaining $\bt h_{a} = \mathcal{E}_{\scriptsize \sfeat}\left[ \diag{ J_g(\sfeat)^\top\bt S^\top \bt S \bt J_g(\sfeat)} \right]$. We compute $\bt h_a$ as the average of the importance maps of different inputs. We call this method frozen WA-RDO or FWA-RDO. This strategy reduces the computational complexity of acquiring the importance map, which is done during training and then kept fixed during inference. Moreover, it removes the need to store the restoration network in the encoder. 

\begin{table}
\setlength{\tabcolsep}{3.15pt}
\renewcommand{\arraystretch}{0.9}
\begin{tabular}{@{}ll c cc c ccc c cc}
\toprule
& & & \multicolumn{2}{c}{\textbf{OIv6} \cite{kuznetsova2020open}} & & \multicolumn{3}{c}{\textbf{SFU HW} \cite{choi2021dataset}} & & \multicolumn{2}{c}{\textbf{Track} \cite{gao2022open, lin2020human}} \\
\cmidrule(lr){4-5} \cmidrule(lr){7-9} \cmidrule(lr){11-12} 
\multicolumn{2}{@{}l}{\textbf{Codec}} & & \textbf{det} & \textbf{seg} & & \textbf{A/B} & \textbf{C} & \textbf{D} & & \textbf{TVD} & \textbf{HE} \\
\midrule
\midrule
\multicolumn{2}{@{}l}{\textbf{Base. vs RI}} & & 7.96 & 11.18 & & 0.99 & 6.82 & 6.80 & & 7.33 & 7.66 \\
\midrule
\multirow{2}{*}{\rotatebox{90}{\textbf{SSE}}} 
& VVC  & & 0.62 & 0.66 & & 0.37 & 0.47 & 0.68 & & 0.76 & 1.71  \\
& HEVC & & 0.20 & 0.24 & & 0.33 & 0.30 & 0.38 & & 0.17 & 0.83  \\
\midrule

\multirow{2}{*}{\rotatebox{90}{\textbf{WA}}}
& AVC  & & 0.21 & 0.16 & & 0.26 & 0.32 & $\textcolor{customblue}{\mathbf{0.80}}$ & & $\textcolor{customblue}{\mathbf{1.21}}$ & 0.42 \\
& HEVC & & 0.34 & 0.54 & & $\textcolor{customblue}{\mathbf{0.41}}$ & $\textcolor{customblue}{\mathbf{0.75}}$ & $\textcolor{customblue}{\mathbf{1.13}}$ & & $\textcolor{customblue}{\mathbf{1.69}}$ & 1.34  \\
\midrule

\multirow{2}{*}{\rotatebox{90}{\textbf{IWA}}}
& AVC  & & – & – & & 0.30 & 0.36 & 0.65 & & $\textcolor{customblue}{\mathbf{1.21}}$ & 0.43  \\
& HEVC & & – & – & & $\textcolor{customblue}{\mathbf{0.40}}$ & 0.46 & $\textcolor{customblue}{\mathbf{1.31}}$ & & $\textcolor{customblue}{\mathbf{1.69}}$ & 1.38  \\
\midrule

\multirow{2}{*}{\rotatebox{90}{\textbf{FWA}}}
& AVC  & & 0.19 & 0.13 & & 0.30 & 0.36 & 0.59 & & $\textcolor{customblue}{\mathbf{0.92}}$ & 0.30 \\
& HEVC & & 0.51 & 0.39 & & $\textcolor{customblue}{\mathbf{0.41}}$ & $\textcolor{customblue}{\mathbf{0.51}}$ & $\textcolor{customblue}{\mathbf{0.86}}$ & & $\textcolor{customblue}{\mathbf{1.59}}$ & 1.26 \\
\bottomrule
\end{tabular}
\caption{BD-accuracy (\%) against FCTM-AVC using SSE-RDO, and baseline BD-accuracy vs (VVC) remote inference (RI). Higher is better. HE stands for HiEve. Since OIv6 contains only images, IWA-RDO cannot be applied (–). For each dataset, we highlight in blue the values that outperform FCTM-VVC with SSE-RDO.}
    \label{tab:results}
\end{table}

\begin{figure*}[th]
    \centering
    \includegraphics[width=\linewidth]{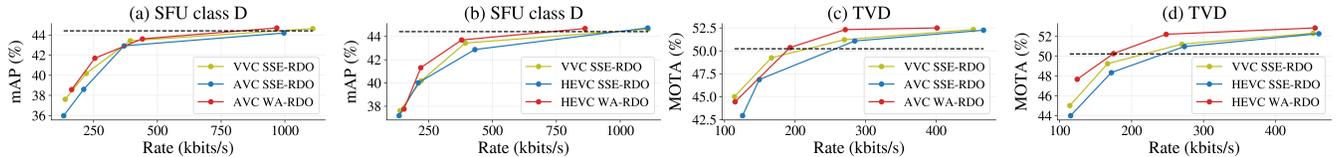}
    \caption{RD curves for (a-b) class D of SFU \cite{choi2021dataset} and (c-d) TVD \cite{gao2022open}, using FCTM-HEVC and FCTM-AVC with WA-RDO and SSE-RDO. We also show the FCTM-VVC anchor, remote inference using VVC, and local inference (dashed line). Our methods improve over SSE-RDO.}
    \label{fig:rd_curves}
    \vspace{-1em}
\end{figure*}

\section{Experiments}
\label{sec:exper}
We consider the test model used by MPEG FCM group (FCTM) \cite{WG04_N0706_2025}, with the four image and video datasets defined in the common test conditions (CTC) \cite{compressai_vision}, for object detection, image segmentation, and object tracking tasks (three wrappers) \cite{eimon2025efficient}. The anchor is VVC FCTM v7.0 (using SSE-RDO). 
We modify AVC (JM 19.1 \cite{wiegand_overview_2003}) and HEVC (HM 18.0), and use these as inner codecs. 
As required by CTC \cite{compressai_vision}, we run our experiments on a CPU (Intel(R) Xeon(R) Platinum 8462Y, single thread, 4.1 GHz). As a GPU, we use an Nvidia(R) RTX 2080.
We report BD-accuracy (BD-SNR with accuracy as distortion) \cite{bjontegaard_calculation_2001}: mean average precision (mAP) for object detection and instance segmentation, and Multiple Object Tracking Accuracy (MOTA) for object tracking. To minimize  crossings, we use the method with worse performance as a baseline (FCTM using SSE-RDO AVC as inner codec).
We also report the BD-accuracy of this baseline against VVC-based remote inference, i.e., compressing and transmitting the video and then running the task on the receiver.

\begin{figure}
    \centering
    \includegraphics[width=\linewidth]{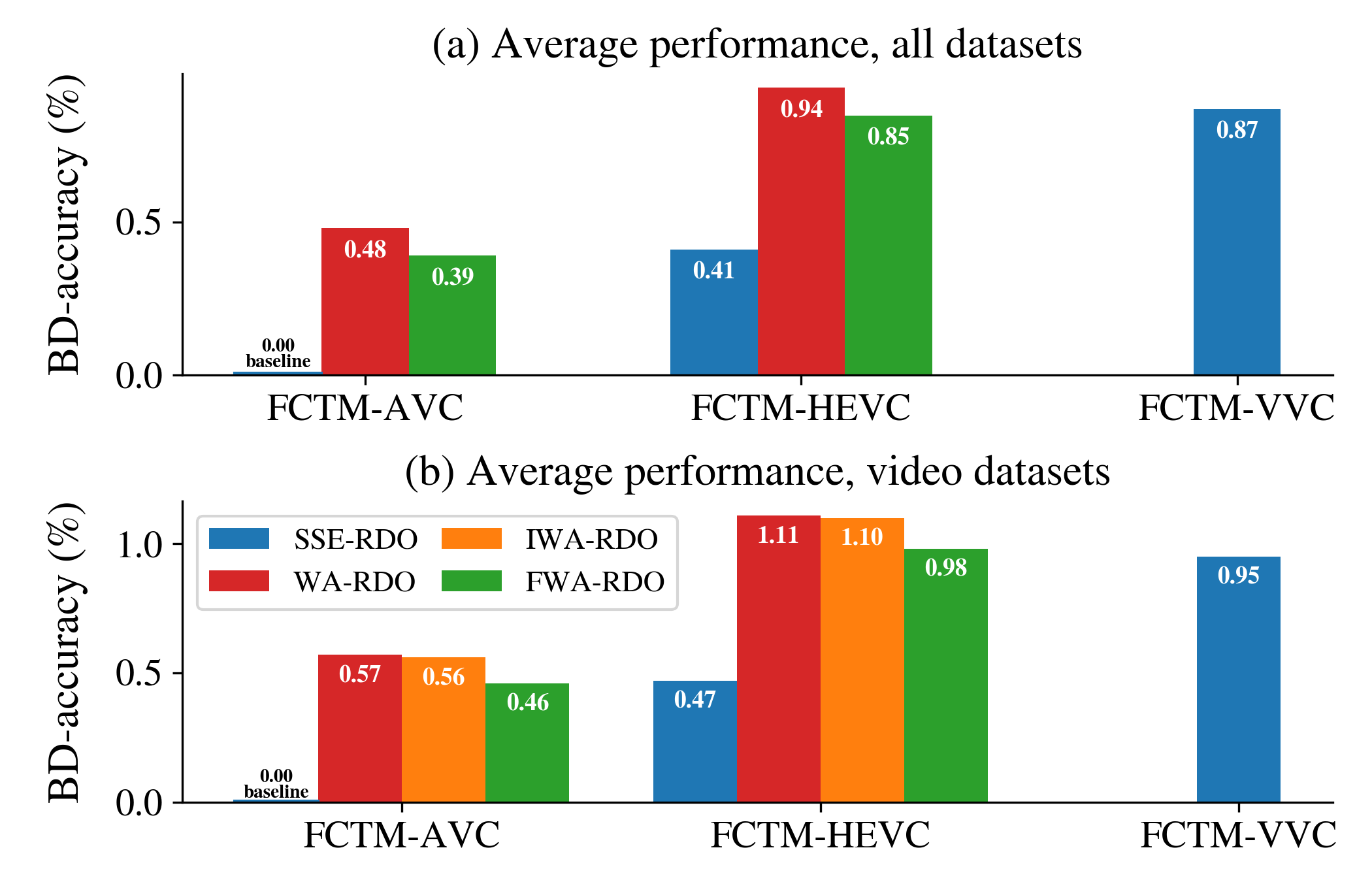}
    \caption{Average BD-accuracy (\%) against FCTM-AVC using SSE-RDO across (a) all the datasets in our experimental setup, and (b) only the video datasets. WA-RDO with HEVC achieves the same rate–accuracy trade-off as the FCTM-VVC anchor. WA-RDO with AVC matches the performance of SSE-RDO FCTM-HEVC.}
    \label{fig:average}
    \vspace{-1em}
\end{figure}

We modify AVC (JM 19.1 \cite{wiegand_overview_2003}, high profile) and HEVC (HM 18.0 \cite{sullivan_overview_2012}, main-RExt profile) so that the block partitioning algorithm and RDOQ are based on our RDO methods. For FWA-RDO, we average the importance maps of $3000$ images from the OIv6 dataset \cite{kuznetsova2020open}, different from those used for validation. 
We test WA-RDO in SFU to assess the role of the sketching dimension used to obtain the importance map in runtime (CPU only and CPU+GPU) and accuracy (\cref{tab:sketch}). We also validated our assumption to obtain \eqref{eq:og_problem} from \eqref{eq:featpres}, evaluating the distortion using SSE-RDO AVC as the inner codec in SFU dataset videos \cite{choi2021dataset} with QPs between $20$ and $35$. Results show that the first term in \eqref{eq:distortion} is, on average, $28.13$ times larger than the last term under the FCTM operating conditions, with the ratio between the maximum of the first term and the minimum of the last term being $14.38$.

\begin{table}[t]
    \centering
    \setlength{\tabcolsep}{4pt} 
    \renewcommand{\arraystretch}{0.6}
    \begin{tabular}{@{}lccccc}
        \toprule
         \textbf{Sketching dim.} & $\mathbf{2}$ & $\mathbf{4}$ & $\mathbf{8}$ & $\mathbf{16}$ &  $\mathbf{32}$  \\
        \midrule
        
            BD-mAP (\%) & $0.26$ & $0.43$ & $0.44$ & $0.47$ & $0.46$ \\
            \midrule
			CPU runtime (s) & $5.92$ & $9.07$ & $15.03$ & $27.76$ & $51.36$ \\
			CPU+GPU runtime (s) & $1.36$ & $1.52$ & $1.88$ & $2.50$ & $3.74$\\            
        \bottomrule
    \end{tabular}    
    \caption{BD-mAP and encoding time per frame (importance map computed on CPU or GPU) for SFU \cite{choi2021dataset} vs sketching samples, for WA-RDO FCTM-AVC. Based on this result, we choose $4$ sketching samples for our other experiments.}
    \label{tab:sketch}
\end{table}	

\smallskip 

\noindent{\textbf{Coding performance.}} We test our methods and SSE-RDO (\cref{tab:results}), showing the average across datasets in \cref{fig:average}, and RD curves in \cref{fig:rd_curves}. Any of our RDO methods outperforms its SSE-RDO counterpart in all datasets and codecs. For datasets where temporal or spatial structures are simple, such as HiEve or OIv6, using a better codec yields better gains than using our method. IWA-RDO reaches similar performance to WA-RDO. FWA-RDO with HEVC can match the performance of VVC-FCTM with SSE-RDO.

\begin{figure}[t]
    \centering
    \includegraphics[width=\linewidth]{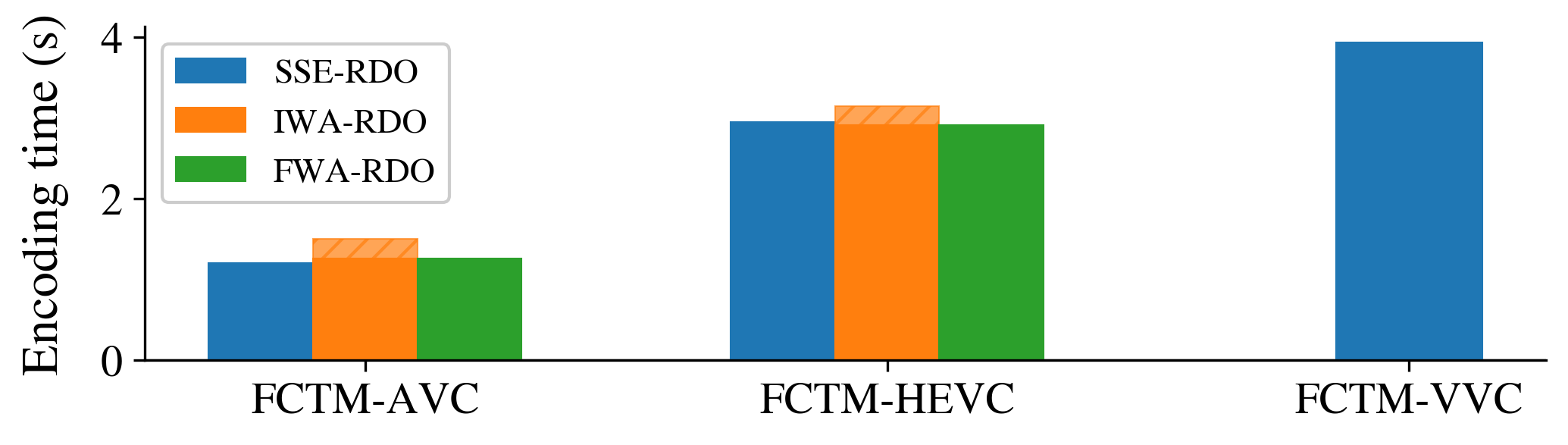}
    \caption{Encoding time per frame (CPU only) for videos in SFU \cite{choi2021dataset} with respect to the FCM anchor, for both AVC and HEVC using SSE-RDO and our methods. Runtimes include importance map computation with $4$ sketching samples, shown with a grid. IWA-RDO and FWA-RDO have similar runtime to SSE-RDO.}
    \label{fig:comp_complex}
    \vspace{-1em}
\end{figure}

\smallskip
\noindent{\textbf{Computational complexity.}} We measure the runtime of the encoder, and compare our methods with SSE-RDO for both AVC and HEVC in \cref{fig:comp_complex}. Both IWA-RDO and FWA-RDO yield similar runtimes to SSE-RDO, with reductions in complexity with respect to VVC of $72\%$ and $28.13\%$ for our modified versions of AVC and HEVC, respectively. FWA-RDO is faster than SSE-RDO in HEVC because the average QP used to obtain the RD curves is larger.  

\section{Conclusion}
\label{sec:majhead}
In this paper, we presented an RDO method for sandwich-based FCM, which accounts for the restoration wrapper during bit-allocation in the inner encoder. We first proposed an RDO method that relies on a wrapper-aware weighted SSE distortion metric. Then, we proposed two simplifications to make it practical with existing FCM codecs: GOP-based temporal reuse, and architecture- and task-based importance maps. Experimental results on the MPEG FCM common test conditions show that our method consistently outperforms conventional SSE-RDO. Most importantly, WA-RDO with an HEVC inner codec matches the task accuracy of the VVC anchor, effectively closing a full codec generation gap at negligible runtime overhead. Future extensions may explore other wrapper-based pipelines \cite{guleryuz_sandwiched_2021}, other codecs \cite{bross2021overview}, and transform design \cite{fernandez_image_2023, fernandez2025int}.

\bibliographystyle{IEEEbib}
\bibliography{ICASSP_2026}
\end{document}

%% file: notation.tex
\newcommand{\diag}[1]{\mathrm{diag}\l \bt #1\r}

\DeclareMathOperator*{\argmin}{arg\,min}

\newcommand{\bt}[1]{\mbox{$\bf #1$}}
\def\l{\left(}
\def\r{\right)}
\newcommand{\img}{\bt x}

\newcommand{\jacos}{\bt J_{\mathbf{s}}}

\newcommand{\numfeat}{n_{\sf f}}
\newcommand{\nummb}{n_b}

\newcommand{\thetavec}{\pmb{\theta}}

\newcommand{\feat}{\bt y}
\newcommand{\cfeat}{\hat{\bt y}}
\newcommand{\sfeat}{\bt z}
\newcommand{\csfeat}{\hat{\bt z}}